\documentclass[nofootinbib,preprint,showpacs,tightenlines]{revtex4-1}

\usepackage[utf8]{inputenc}
\usepackage{amsmath}
\usepackage{amssymb}
\usepackage{graphicx}

\newcommand{\vierint}{ \int_{|\mathbf{q}|<\Lambda}   \frac{d^4q}{\left( 2 \pi \right)^4}}
\newcommand{\dreiint}{ \int_{|\mathbf{q}|<\Lambda}  \frac{d^3q}{\left( 2 \pi \right)^3}}
\newcommand{\einhalb}{ \tfrac{1}{2}}

\begin{document}

\title{Pauli blocking effects and Cooper triples in three-component Fermi gases}

\author{P. Niemann}
\email{niemann@hiskp.uni-bonn.de}
\author{H.-W. Hammer}
\email{hammer@hiskp.uni-bonn.de}
\affiliation{Helmholtz-Institut f\"ur Strahlen- und Kernphysik (Theorie)\\
and Bethe Center for Theoretical Physics, Universit\"at Bonn, 53115 Bonn,
Germany}

\date{\today}

\begin{abstract}
We investigate the effect of Pauli blocking on universal two- and 
three-body states in the medium. Their corresponding energies are 
extracted from the poles of two- and three-body in-medium scattering 
amplitudes.  Compared to the vacuum, the binding of dimer and trimer states 
is reduced by the medium effects. In two-body scattering,
the well-known physics of Cooper pairs is recovered.
In the three-body sector, we find a new class of positive
energy poles which can be interpreted as Cooper triples.
\end{abstract}

\maketitle

\section{Introduction}

Ultracold Fermi gases offer a unique possibility to investigate many-body 
phenomena in a controlled environment
\cite{Giorgini:2008,Ketterle:2008,Lee:2008fa,Chin:2010aa}. 
In dilute systems of two-component
fermions, the interactions are characterized by the S-wave scattering length.
Close to a Feshbach resonance, the scattering length 
can be tuned experimentally by varying an external magnetic field.  
In particular, the crossover from the
BCS limit of weakly interacting fermions to a BEC of bosonic dimers 
by tuning through a resonance has been studied in great detail
\cite{Chin:2010aa}.
The behavior of such a system is constrained by universal relations 
that involve the so-called contact, which measures the number of pairs of 
fermions with different spins that have small separations 
\cite{Tan:2005xx,Braaten:2010if}.

More recently, ultracold gases of three-component fermions
have also been investigated. The interest in such systems has various 
motivations.
First, the manifestation of the Efimov effect \cite{Efimov:1970zz}
has been studied
in systems consisting of three hyperfine states of fermionic
$^6$Li atoms. A resonant  enhancement of the recombination rates
at certain values of the scattering lengths was observed
in experiment \cite{Ottenstein:2008,Huckans:2008}.
These observations were analyzed theoretically and
traced back to the appearance of
an Efimov trimer close to the three-atom threshold
\cite{Braaten:2008wd,schmidt:2008fz,Naidon:2009,Braaten:2009ey}.
Subsequently, the direct 
association of Efimov trimers was also achieved~\cite{Lompe:10,Nakajima:11}. 

A second line of research has focused on the 
phase structure of such systems
\cite{pair-three-comp,super-phases,Breached-Pairing,Catelani:2008}. 
In these theoretical studies, two components are typically paired
while the third one remains unpaired.
This mechanism can be 
regarded as a generalization of the BCS case. Moreover, the BEC-BCS 
crossover has also been investigated in a three component system.
In Ref.~\cite{Crossover-3komp}, the dynamics of such a system 
was analyzed on time scales long enough
to see two-body physics but short enough to be able to
neglect Efimov states or three-body collisions.
For three-component fermions in an optical lattice, the formation of a superfluid phase at weak coupling
and a \lq\lq trion'' phase of three-fermion 
bound states at strong coupling has been predicted
\cite{Rapp:2006rx}.

In this work, we combine both lines of research and investigate
three-body correlations in the medium.
We investigate the effect of Pauli blocking induced by the presence
of a Fermi sphere on universal two- and 
three-body states in the medium. Their corresponding energies are 
extracted from the poles of two- and three-body scattering amplitudes
in the medium. 
A similar study was carried out in Ref.~\cite{P-Block-Ef} for the case
of a fermion immersed in a Fermi sea interacting with two heavy bosons. 
The Born-Oppenheimer approximation was used to map the system 
to an effective two-body problem and calculate the dependence of the 
universal  spectrum of Efimov trimers on the Fermi density in that case. 
In Ref.~\cite{Nygaard:2011aa}, the modification of the Efimov
spectrum for three equal-mass fermions when one of the fermions
is embedded in a Fermi sea was calculated numerically and the 
modification of the universal scaling 
behavior by the background density of fermionic particles was 
investigated. 

Here, we investigate the medium 
modifications for three equal-mass fermions  all of which are 
embedded in a Fermi sphere. We solve the two- and
three-body scattering equations 
for this system (cf. Ref.~\cite{Braaten:2008wd}) in the medium and present
a detailed study of the poles of the in-medium scattering amplitude.
In particular, we study the emergence of positive energy 
three-body poles analog to
the Cooper pairs in the two-body system.
A similar analysis was carried out in Refs.~\cite{Schuck-BF} for the
in-medium scattering amplitude of a boson and a fermion. In these
studies, the boson-fermion Cooper pairs were found to persist for
vanishing attraction.

We consider three distinguishable 
non-relativistic particles of equal mass with 
resonant interactions in a Fermi sea at zero temperature.
The system is assumed to be dilute, i.e. $k_F R\ll 1$, where $R$
is the range of the interaction and $k_F$ the Fermi momentum. 
In this case, the two-body interactions of the particles are determined  
by their scattering length $a$. We assume the two-body interactions to be 
resonant, i.e. $|a| \gg R$.
Effective range corrections are suppressed and can be treated
in perturbation theory. Because we are at zero temperature, all states up to 
$k_F$ are occupied. For $k_F a \ll 1$, a perturbative low-density expansion
can be derived~\cite{Hammer:2000xg}, but for $k_F a \sim 1$ an infinite
class of diagrams has to be resummed and one has to resort to Monte Carlo 
simulations or additional expansions~\cite{Lee:2008fa,Furnstahl:2008df}.
In this study, we include only the interaction of the particles 
with the Fermi sea via Pauli blocking. These effects dominate in an 
expansion in the inverse number of dimensions  
\cite{Steele:2000qt,Schafer:2005kg} and determine the qualitative 
behavior of the system.  Other effects of the medium,
such as the excitation of particles out of the Fermi sea through
scattering processes, are neglected. 

Our theoretical framework is based on
an effective Lagrangian for the fermion fields
$\Psi_i$, $i=0,1,2$:
\begin{equation}
\mathcal{L} = \sum_{i=0}^2 \Psi_i^{\dagger}\left( i \partial_t + 
\frac{ \vec{\nabla}^2 }{2m}\right)\Psi_i -\sum_{k=0}^{2} \frac{g_{k}}{2} 
\Psi_i^{\dagger} \Psi_j^{\dagger} \Psi_i \Psi_j +  h\, \Psi_0^{\dagger}\Psi_1^{\dagger}\Psi_2^{\dagger}\Psi_0\Psi_1\Psi_2\:,
\label{Lagrangedichte1}
\end{equation}
where the coupling constant ${g_{k}}$ with $i\neq j\neq k$ parametrizes the 
interaction of fermions $i$ and $j$. 
The term proportional to $h$ is a contact
three-body interaction of all three fermions. It determines
the spectrum of three-body Efimov states in the vacuum.
The explicit form of this
term will not be required for our study, since the dependence on the 
three-body term can be traded for a dependence on the cutoff in leading order
calculations  \cite{Hammer:2000nf}. For
practical calculations, it is convenient to introduce auxiliary dimer 
fields $d_k$ and rewrite the Lagrangian in the form:
\begin{equation} 
\mathcal{L} = \sum_{i=0}^2 \Psi_i^{\dagger}\left( i \partial_t + 
\frac{ \vec{\nabla}^2 }{2m}\right)\Psi_i +\sum_{k=0}^2 \left( 
\Delta_k d_k^{\dagger} d_k -\frac{g_k}{2} \left( d_k^{\dagger} 
\Psi_i \Psi_j + \Psi_i^{\dagger} \Psi_j^{\dagger} d_k \right) \right) 
+   h\, \Psi_0^{\dagger}\Psi_1^{\dagger}\Psi_2^{\dagger}\Psi_0\Psi_1\Psi_2\:.
\label{Lagrangedichte2}
\end{equation}
The dimer field $d_k$ describes two interacting particles $i$ and $j$ with 
$i \neq j \neq k$. Using the classical equations of motion, the equivalence 
of equation (\ref{Lagrangedichte1}) 
and (\ref{Lagrangedichte2}) can be demonstrated. This framework has been 
widely used to describe the universal properties of few-body systems close
to the universal limit \cite{hw-review}. It has also been used as
the basis for studies of the Efimov effect in systems three-component fermions
\cite{Braaten:2008wd,Braaten:2009ey}.

\section{Two-body sector}

\subsection{Vacuum case}
\begin{figure}[ht]
\centering
\includegraphics[clip,width=15cm,angle=0]{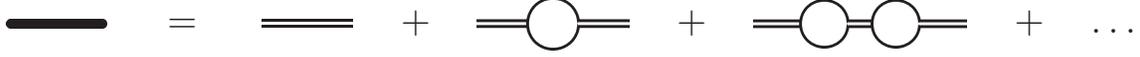}
  \caption{Bubble sum for the full interacting dimer propagator (thick line).
  The double lines correspond to the bare dimer propagator and the single 
  lines indicate particle propagators.}
  \label{VollerDimer}
\end{figure}

We are now in the position to investigate the effect of Pauli blocking 
on universal two- and 
three-body states in the medium.
We start by briefly reviewing the vacuum case. 
More details can be  found in Ref.~\cite{hw-review}.
For convenience, we set $\hbar=m=1$ from now on.
The bare dimer propagator derived from the Lagrangian (\ref{Lagrangedichte2})
is simply a constant, $i/\Delta_k$.
The full, interacting dimer propagator is given by dressing the bare
propagator with fermion bubbles, see Fig.~\ref{VollerDimer}.
It represents the exact solution of the vacuum two-body problem
for the Lagrangian (\ref{Lagrangedichte2}).
The diagrams constitute a geometric series, 
which can easily be summed. The result can be written as
\begin{align}
iD_k(P_0,\mathbf{P}) = \frac{i}{\Delta_k\left(1-\frac{g_k^2}{4\Delta_k}I(P_0,\mathbf{P}) \right)} \:,
\end{align}
where $I(P_0,\mathbf{P})$ is the loop function for 
the two-fermion bubble in Fig.~\ref{VollerDimer}.
In the vacuum, the loop function is
\begin{eqnarray}
iI(P_0,\mathbf{P})  &=& \vierint \frac{i }{\frac{P_0}{2}+q_0 -\frac{1}{2}\left( \frac{\mathbf{P}}{2} +\mathbf{q} \right)^2 +i\epsilon}    
\frac{i }{\frac{P_0}{2}-q_0 -\frac{1}{2}\left( \frac{\mathbf{P}}{2} -\mathbf{q} \right)^2 +i \epsilon} \nonumber\\
&=& \frac{i}{4\pi}\left( -\frac{2\Lambda}{\pi} +\sqrt{-P_0+P^2/4-i \epsilon}\right)\:,
\label{Loop-Vak}
\end{eqnarray}
where $P \equiv |\mathbf{P}|$ and
the UV divergence of the loop integral 
has been regulated by a momentum cutoff $\Lambda$. The cutoff dependence is
absorbed into the coupling constant $g_k$, 
such that all observable quantities are independent of $\Lambda$. 
The two-body scattering amplitude is obtained by multiplying the 
full dimer propagator with the square of the dimer-fermion coupling, $(-ig_k/2)^2$.
Matching to the amplitude for scattering of particles $i$ and $j$ 
in the center of mass at energy $E=p^2$,
\begin{equation}
T_k(p^2)= \frac{4\pi}{-1/a_k -ip} \stackrel{!}{=} -\frac{g_k^2}{4} 
D_k(p^2,0)\:,
\end{equation}
we obtain
\begin{align}
&\frac{ g_k^2} {\Delta_k}=\frac{16\pi a_k}{1-2a_k\Lambda/\pi} \:.
 \label{ak-matching}
\end{align}
Note that $ g_k$ and $\Delta_k$ are not independent at this order
and all observables depend on the combination $g_k^2/\Delta_k$. The renormalized dimer propagator 
in the vacuum can thus be written as
\begin{equation} 
iD_k(P_0,\mathbf{P})_{vak}=i \frac{16\pi}{g_k^2} \left[ 1/a_k -\sqrt{-P_0+P^2/4-i \epsilon} \right]^{-1} \: .
\label{eq:vacprop}
\end{equation}
The propagator has a bound state pole at $P_0= -1/a_k^2 + P^2/4$ if $a_k>0$.
The energy at the pole is composed of the binding energy $-1/a_k^2$  
and the kinetic energy of the dimer $P^2/4$. The total 
mass is $2m=2$, as expected for a dimer state. For negative scattering length, the pole is on the 
unphysical sheet and represents a virtual state.

\subsection{Medium case}
We now move on to medium case.  
In the presence of a Fermi sphere, the loop integral changes to
\begin{equation}
iI(P_0,\mathbf{P})  = \vierint \frac{i\Theta \left(| \frac{ \mathbf{P}}{2}  +\mathbf{q}| 
-k_F \right) }{\frac{P_0}{2}+q_0 -\frac{1}{2}\left( \frac{\mathbf{P}}{2} +\mathbf{q} \right)^2 +i\epsilon}    
\frac{i\Theta \left(| \frac{ \mathbf{P}}{2}  -\mathbf{q}| -k_F \right) }{\frac{P_0}{2}-q_0 
-\frac{1}{2}\left( \frac{\mathbf{P}}{2} -\mathbf{q} \right)^2 +i \epsilon} \:,
\end{equation}
where the theta functions encode the Pauli blocking.
They ensure that the intermediate particles can not scatter into occupied 
states in the Fermi sea.  This introduces boundary conditions for the loop integrals at small 
momenta.  Different cases must be considered. A summary of the calculation 
is given in Appendix~\ref{sec:medium-integ}. Here we focus on the results.

For vanishing total momentum $P$ the boundary conditions become simple. In this case, the argument of both 
theta functions is $|\mathbf{q}|-k_F$. Consequently, the integration over $|\mathbf{q}|$
starts at $k_F$ and ends at $\Lambda$. It is evident that only the infrared behavior of the integrals
is modified by the Fermi sea. The renormalization of UV divergences is the same as in the vacuum.
The in-medium dimer propagator then has the form 
\begin{align}
iD_k(P_0,P)=i  \frac{16\pi}{g_k^2}\left[ \frac{1}{a_k} -\frac{1}{\pi}L(P_0,P)  \right]^{-1} \: ,
\end{align}
with 
\begin{equation}
L(P_0,P=0)=2k_F + \sqrt{P_0+ i \epsilon} \left[ \ln(k_F-\sqrt{P_0+ i\epsilon})
-\ln(k_F+\sqrt{P_0+ i \epsilon}) \right] \: .
\end{equation}
The poles of the propagator are determined by solving
\begin{equation}
\frac{1}{a_k}= \frac{2k_F}{\pi} + \frac{\sqrt{P_0+ i \epsilon}}{\pi} \left( \ln 
\left( k_F-\sqrt{P_0+i \epsilon}\right)-\ln \left(k_F+\sqrt{P_0+ i \epsilon} \right) \right) \: 
\end{equation}
for $P_0$. If $P_0$ is negative, this equation can be written as
\begin{equation}
\frac{1}{a_k}=\frac{2k_F}{\pi}+\frac{2}{\pi} \sqrt{|P_0|}\arctan \left( \frac{\sqrt{|P_0|}}{k_F} \right) \:,
\end{equation}
where the  $i \epsilon$ has been omitted.
This equation is formally similar to Eq.~(3) of Ref.~\cite{P-Block-Ef} for the binding energy of a light fermion  
immersed in a Fermi sea interacting with two heavy bosons. In this case, the Born-Oppenheimer approximation
may be used and the three-body problem reduces to an effective two-body problem.

In the general case, the boundary conditions are more complex (cf. Appendix \ref{sec:medium-integ}). 
Two cases have to be distinguished: 
$P<2k_F$ and $P > 2k_F$. The result for the general in-medium loop function $L(P_0,P)$ is: 

\begin{itemize}
\item[(a)] $P<2k_F$:
\begin{align}
L(P_0,P)=&P/2+k_F+\sqrt{\sigma}\left[ \ln\left(P/2+k_F-\sqrt{\sigma}\right) 
-\ln \left(P/2+k_F+\sqrt{\sigma}\right) \right] \notag\\
 &+\frac{k_F^2-P_0 -i \epsilon}{P} \bigg [ \ln \left( P/2+k_F-\sqrt{\sigma}\right)+
 \ln \left( P/2+k_F+\sqrt{\sigma} \right) \notag \\
 & - \ln \left( \sqrt{k_F^2-\tfrac{1}{4}P^2}-\sqrt{\sigma} \right) - 
\ln \left( \sqrt{k_F^2-\tfrac{1}{4} P^2}+\sqrt{\sigma} \right) \bigg ]\,,
\label{eq:medprop1} 
\end{align}
\item[(b)] $P>2k_F$:
\begin{align}
L(P_0,P)=&2k_F   + \pi \sqrt{-\sigma} 
+ \sqrt{\sigma} \bigg[\ln \left(P/2-k_F+\sqrt{\sigma} \right) 
+ \ln \left( P/2 +k_F-\sqrt{\sigma}\right) \notag\\
&-\ln \left( P/2-k_F-\sqrt{\sigma}\right) 
- \ln \left( P/2 +k_F+\sqrt{\sigma} \right) \bigg ] \notag \\
&+\frac{-k_F^2+P_0+i\epsilon}{P} \bigg[ \ln \left( P/2-k_F-\sqrt{\sigma} \right) 
+ \ln \left( P/2 -k_F+\sqrt{\sigma}\right)  \notag \\
&- \ln \left(P/2+k_F-\sqrt{\sigma} \right) - \ln \left( P/2 + k_F + \sqrt{\sigma} \right) \bigg ]\,,
\label{eq:medprop2} 
\end{align}
\end{itemize}
with $\sqrt{\sigma}=\sqrt{P_0-P^2/4 + i \epsilon}$\:.

We now discuss the poles of the dimer propagator in the medium. Our aim is to recover the known two-body physics
from the viewpoint of the pole structure and then use the same strategy to understand the three-body sector.
First, we specify our units. Since there is one free length scale $l_0$ in the calculations, 
we express all dimensionful quantities in units of $l_0$: the energy has the unit $[1/l_0^2]$, 
scattering lengths $[l_0]$ and momenta $[1/l_0]$.

\begin{figure}[t]
\centering
\includegraphics[clip,width=10cm,angle=0]{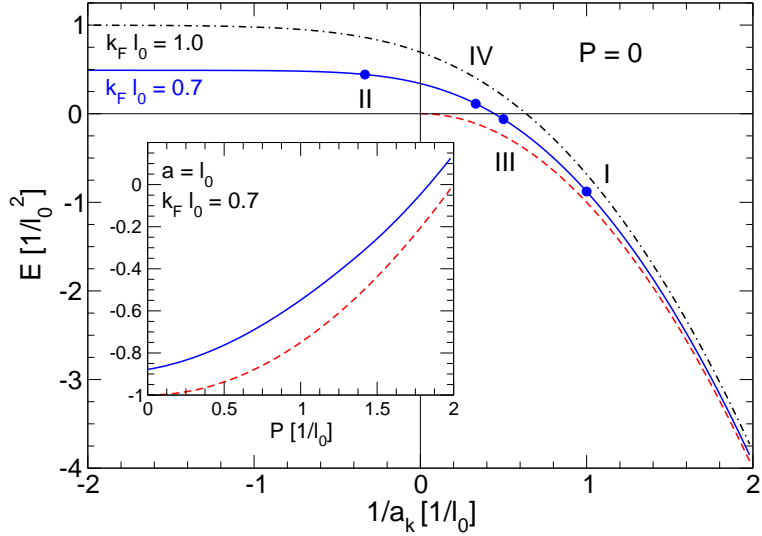}
\caption{(Color online) The energy $E$ of the dimer pole at $P=0$
plotted against the inverse scattering length 
$1/a_k$ for a Fermi momentum $k_F=0.7/l_0$ (solid line) and $k_F=1/l_0$ (dash-dotted line). 
In addition, four selected points are marked I, II, III, and IV
on the solid line. For comparison, the vacuum pole
energy is shown by the dashed line. In the inset, the dimer pole energy  is displayed 
as a function of the total momentrum $P$ for $k_F=0.7/l_0$ and $a=l_0$. Curves are as 
above.  
}
\label{fig:Streulaenge1}
\end{figure}
We never find more than one pole on the physical sheet in the in-medium 
dimer propagator. The physical conditions under which this pole can 
disappear are discussed below.
In Fig.~\ref{fig:Streulaenge1}, the energy of the pole, $E$, is plotted against the inverse 
scattering length $1/a_k$ at vanishing momentum $P=0$ for $k_F l_0 =0.7$ (solid line)
and $1$ (dash-dotted line), respectively. The dashed curve represents 
the dimer energy in the vacuum case. For positive scattering length, the energy of the vacuum 
pole is $ E=-1/a_k^2$.  There is no vacuum pole on the physical
sheet if the scattering length is negative. 
For non-vanishing Fermi momentum, a pole with positive energy appears in the negative 
scattering length region.  In the limit $1/a_k \to -\infty$, this pole asymptotically approaches 
the values $(k_F l_0)^2=0.49$ and $1$ for $k_F l_0 =0.7$ and $1$, respectively. 
In the positive scattering length region, the pole behaves like a vacuum pole if the 
scattering length is sufficiently small. However, the corresponding binding energy is reduced
by medium effects. Additionally four selected points are marked on the solid line: I, II, III, and IV. 
To gain deeper insight into the nature of the pole in the in-medium dimer propagator, 
these points will be further investigated below. 
The parameters $k_F$ and $a$ are kept fixed while the momentum $P$ will be varied.

In the inset of Fig.~\ref{fig:Streulaenge1}, which corresponds to point I, the dependence of 
the pole energy on the total momentum $P$ is shown
for $a_k=l_0$ and $ k_F=0.7/l_0$ (solid line). The dashed line shows the vacuum pole energy as before.
Medium and vacuum poles have a similar behavior as function of the total momentum: with increasing 
momentum $P$, the pole energy is increased and the binding is reduced. 
For the vaccuum pole, this is evident from Eq.~(\ref{eq:vacprop}).
In the medium, it follows from the dominant functional dependence of
the in-medium dimer propagator on $\sigma=P_0-P^2/4$
(cf.~Eqs.~(\ref{eq:medprop1}, \ref{eq:medprop2})). Moreover, medium effects are again seen to lower 
the pole energies. 

\begin{figure}[t]
\centering
\includegraphics[clip,width=12cm,angle=0]{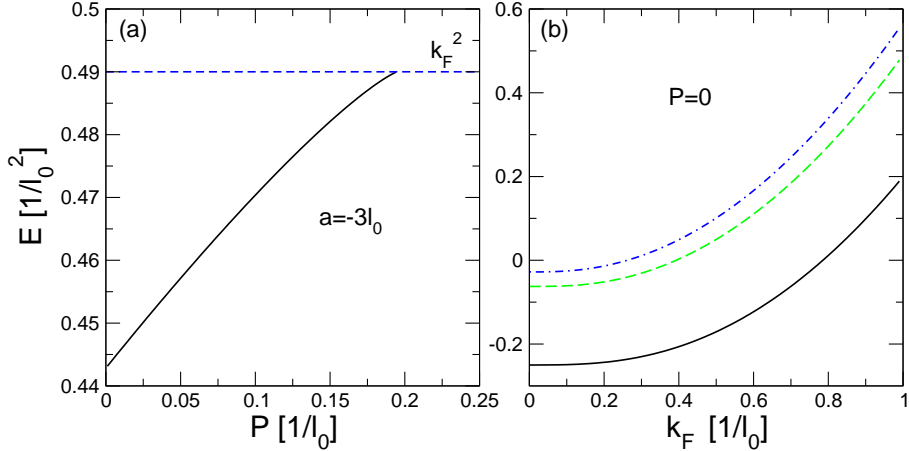}
\caption{(Color online) 
(a) The energy $E$ depicted as function of the total momentum $P$ 
for $a_k=-3l_0$ and $k_F=0.7/l_0$ (solid line). The vertical dotted line 
gives $E=k_F^2$. (b) Energy $E$ plotted against the Fermi momentum 
$k_F$ for $a_k$=$2l_0$ (solid), $4l_0$ (dashed), $6l_0$ (dashed dotted); $P=0$.
}
\label{fig:dimer-kf2}
\end{figure}
Next, we examine the positive energy poles more thoroughly. In Fig. \ref{fig:dimer-kf2}~(a), the energy 
is plotted against the total momentum for a negative scattering length  $a_k=-3 l_0$, 
which corresponds to point II in Fig.~\ref{fig:Streulaenge1}. The energy of 
the pole is positive and continuously rises as the momentum $P$ is increased
until the energy reaches the value $k_F^2 = 0.49/l_0^2$, 
where the pole disappears.
This positive energy pole can be associated with Cooper pairs~\cite{Fetter}. 
With this interpretation, their peculiar behaviour can be understood. Assume that the two particles are inside the Fermi 
sphere. 
If there is no interaction, the energy of the particles is just 
their kinetic energy. In the presence of attractive interactions, the energy of the two particles, given by the 
pole energy, is lowered. 
Consequently, the energy gain $\Delta E$ is the difference of the kinetic energy and the pole energy. Because the 
maximum kinetic energy of two particles inside the Fermi sea is $k_F^2/2 + k_F^2/2$, 
the maximum energy of the pole is also $k_F^2$. 
When the total momentum of the two particles becomes too large,  the pole disappears.
This property is compatible with the intepretation as Cooper pairs, 
whose total momentum is commonly assumed to be zero. Remember that the energy threshold already appeared in 
Fig.~\ref{fig:Streulaenge1}. But in this instance a different limit was considered.
The energy of the pole approaches the threshold $k_F^2$ asymptotically
in the limit $1/a_k \rightarrow  - \infty$. The energy gain $\Delta E$, hence, decreases 
in this limit. But the poles never disappear and Cooper pairs can always be formed in this region. 

We now turn to the dependence of the poles on the Fermi momentum in the positive scattering length region.
In Fig. \ref{fig:dimer-kf2}~(b), the pole energy is plotted against the Fermi momentum. The total momentum $P$
is set to zero and the scattering lengths are $a_k=2 l_0,\, 4 l_0$ and $6 l_0$. As expected, the energy is negative 
at $k_F =0$ and the pole corresponds to a bound state. With increasing Fermi momentum, the medium effects
become stronger and the binding is reduced. For small $k_F$ the energy rises only slowly but at larger
Fermi momentum the energy changes rapidly, crosses zero, and becomes positive. 
Hence, we observe a continuous crossover from bound states to positive energy poles as the Fermi momentum
is increased.

So far, we could associate the left and right regions in Fig.~\ref{fig:Streulaenge1} to Cooper pairs (cf. II) and bound 
states (cf. I), respectively. 
In between lies the crossover region. We will investigate this region further at the two remaining points III and IV.
Here the nature of the poles changes as a function of momentum. 
\begin{figure}[t]
\centering
\includegraphics[clip,width=12cm,angle=0]{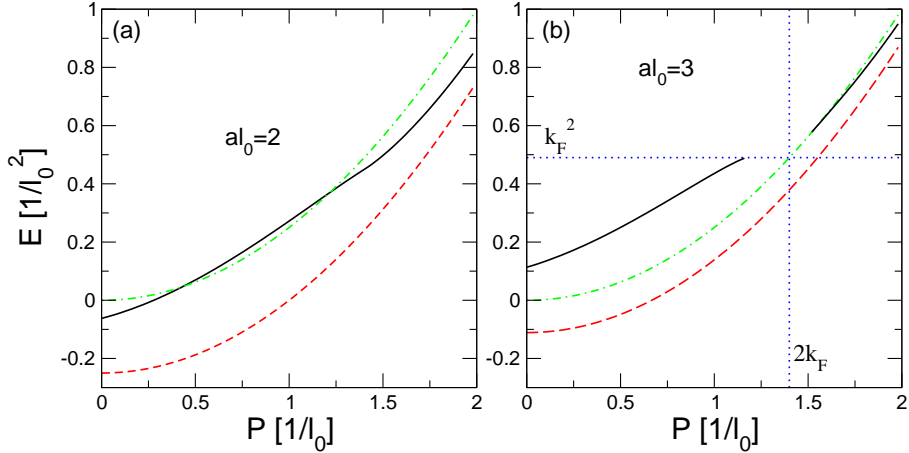}
\caption{(Color online) 
The energy of the poles plotted as a function of the total momentum $P$ for 
$a=2l_0$ [panel (a)] and $a=3l_0$ [panel (b)] and $k_F=0.7l_0$
(solid line). The dashed line shows 
the vacuum poles and the dash-dotted line gives the kinetic energy $P^2/4$. 
The horizontal dotted line in (b) gives $k_F^2$ and the vertical dotted line gives $2k_F$. }
\label{fig:crossover}
\end{figure}
In Fig.~\ref{fig:crossover}~(a) the energy of the pole is  plotted against the total momentum (solid line) with 
$a= 2l_0$ corresponding to point III in  Fig.~\ref{fig:Streulaenge1}.
By comparision, the dashed line shows the vacuum pole and the dash-dotted line the kinetic energy $P^2/4$. 
For vanishing momentum the energy of the pole is extremely reduced compared to the vaccuum
but still negative. However, in the region around 
$P=l_0$ the energy becomes bigger than the kinetic energy. Hence, this pole can not correspond to a bound state
in this region. For larger momenta the energy again drops below the kinetic energy and the poles behave 
similiar to vacuum poles.

We now turn to point IV in the crossover region. In Fig.~\ref{fig:crossover}~(b) an entirely different behaviour can 
be observerd. Again the vacuum pole (dashed line) and the kinetic energy (dash-dotted line) are shown. Striking 
are the three qualitatively different regions in this graph. For momenta $P<2k_F$ the pole seems to correspond to
a Cooper pair: at $P=0$ the energy of the poles is positive, in this whole region the energy is larger than the 
kinetic energy, and the pole disappears when the energy approaches $k_F^2$. In a region around $P\approx 2k_F$ 
there is no pole at all. For slightly larger momenta, the pole reappears. At first the energy is very close 
to the kinetic energy, but for larger momentum it approaches the vacuum energy, as expected. 
These pole now behaves like a bound state.

In summary, we have related the positive energy poles at $P=0$ to Cooper pairs and the negative energy poles to bound 
states. A finite momentum $P$ leads to an increase in the pole energy. In the vacuum, the additional energy is simply
the kinetic energy $P^2/4$. In the medium, the pole energy also increases but the dependence on $P$ is more complicated. 
In particular, the poles can vanish and change their character.
As the Fermi momentum $k_F$ is increased, e.g.,
the binding energy is reduced by medium effects. 
We identified the two extremes I and II in Fig.~\ref{fig:Streulaenge1} with the BCS and BEC domains,
respectively. In between there is a crossover region. In this region the poles change their character 
as a function of the momentum $P$ and they
can not be uniquely related to one of the two cases. Equipped with this qualitative understanding of
in-medium two-body physics, we move on to the three-body amplitude.

\section{Three-body sector}
\subsection{Vaccuum case}
We start by briefly reviewing the physics issues of the vacuum case and then move on
to the medium. In the three-body system with resonant interactions, there is a
universal spectrum of three-body bound states with an accumulation point at zero
energy, called Efimov states \cite{Efimov:1970zz}.
The spectrum is given by Efimov's universal equation
\begin{eqnarray}
E_B^{(n)}  + {1\over a^2} =
\left(e^{-2 \pi/ s_0} \right)^{n-n_*}
\exp \left[ \Delta ( \xi )/s_0 \right] \kappa_*^2 \,,
\label{eq:bind}
\end{eqnarray}
where the angle $\xi$ is defined by
\begin{eqnarray}
\label{xin-def}
\tan \xi = - a\sqrt{E_B^{(n)}} \,,
\end{eqnarray}
$s_0\approx 1.00624$ is a transcendental
number, and $\kappa_*$ is the binding wave number of the state labelled $n_*$.
The function $\Delta ( \xi )$ was first calculated in Ref.~\cite{Braaten:2002sr}
and satisfies $\Delta( -{1\over2}\pi )=0$.
In the unitary limit of infinite scattering length, the spectrum thus becomes 
geometric.
The qualitative features of this spectrum are determined by the scattering length $a$, but
the exact energies depend on the three-body interaction in Eq.~(\ref{Lagrangedichte2})
which fixes the value of $\kappa_*$ \cite{Bedaque:1998kg}.
The spectrum exhibits a discrete scaling symmetry which is evident in Eq.~(\ref{eq:bind}):
if the scattering length $a$ and the energies $E_B$ are rescaled by the discrete
scaling factor $\lambda=\exp(\pi/s_0)$ and $\lambda^{-2}$, respectively, but $\kappa_*$ remains fixed
the spectrum is mapped onto itself. If the scattering length dependence of one state is known, thus all other can be
obtained from the scaling transformation. A detailed discussion 
of these issues can be found in Ref.~\cite{hw-review}. Here, we focus on the modification of
this spectrum in the medium and on possible positive energy poles in the three-body amplitude
similar to the two-body case discussed above.
As discussed above, we set the three-body interaction to zero in our calculation. 
Thus $\kappa_*$ is proportional to the momentum cutoff $\Lambda$. The exact proportionality factor
is not required for our purpose.  
A detailed study of the Efimov spectrum and the universal scaling relations
in the presence of one Fermi sphere was carried out in Ref.~\cite{Nygaard:2011aa}.
We go beyond this study by considering three Fermi spheres and explicitly focusing
on the emergence positive energy poles in the three-body amplitude.
Preliminary results of our study were already presented in~\cite{PatrickDip}.

\subsection{Medium case}
\begin{figure}[t]
\centering
\includegraphics[clip,width=14cm,angle=0]{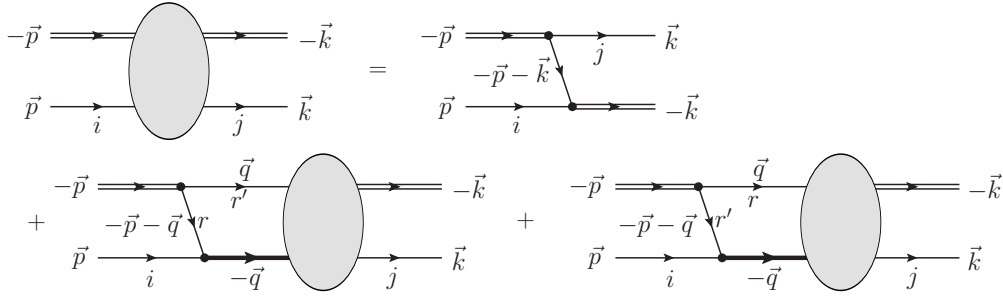}
\caption{Feynman diagrams for the fermion-dimer scattering amplitude for zero
total momentum. Momenta $\mathbf{p},\,\mathbf{q},\,\mathbf{k}$
and fermion indices $i,j,r,r'$ are assigned as in Eq.~(\ref{eq:3bfeyn}).}
\label{Fig:Streuamplitude}
\end{figure}

The three-particle scattering amplitude in the medium 
can be calculated by solving an integral equation. 
In order to simplify the boundary conditions given by the Pauli blocking,  we will constrain 
the total momentum of the three particles to be zero. We note that the Fermi sea provides a special 
reference frame and a non-zero momentum can not be obtained from a simple Galilei 
transformation. However, we have seen in the two-body case that a non-zero
momentum essentially increases the pole energy. Outside of the 
crossover region, the qualitative behavior remains unchanged (cf. inset of Fig.~\ref{fig:Streulaenge1}). 
We expect the same to be true in the three-body case.
The Feynman diagrams for three-body scattering amplitude are depicted in Fig.~\ref{Fig:Streuamplitude}.
Since we are interested in the three-body singularities of the amplitude, it is sufficient to consider the 
fermion-dimer scattering amplitude where the external dimer propagators are amputated.
Because the three particles are distinguishable, 
we have one inhomogeneous and two homogeneous contributions 
to the amplitude as the intermediate dimer can be formed in two ways \cite{Braaten:2008wd}. 
The integral equation for the amplitude $\mathcal{A}_{ij}$
can be written as 
\begin{align}
i\mathcal{A}_{ij}(\mathbf{p},\mathbf{k},E,E_i,E_j) =& -\frac{g_i g_j}{4}  
\frac{i\theta(|\mathbf{p}+\mathbf{k}|-k_F)}{E-E_i-E_j-\frac{(\mathbf{p}+\mathbf{k})^2}{2}
+ i \epsilon} \cdot (1-\delta_{ij}) \notag \\ 
&+ \sum_{r=0}^2 -\frac{g_i g_r}{4}  \vierint \frac{i\theta(q-k_F)}{q_0-\frac{1}{2}q^2+ 
i \epsilon} \cdot \frac{i\theta(|\mathbf{p}+\mathbf{q}|-k_F)}{E-E_i -q_0 - \frac{1}{2}
(\mathbf{p}+\mathbf{q})^2+ i \epsilon}   \notag \\
& \times iD_r (E-q_0,q)\cdot(1-\delta_{ir}) \cdot i \mathcal{A}_{rj}(\mathbf{q},\mathbf{k},E,q_0,E_j) \:,
\label{eq:3bfeyn}
\end{align}
where the momenta and particle indices are assigned as in Fig.~\ref{Fig:Streuamplitude} and $E$ is
the total energy.
After setting the energies  of the incoming and outgoing particles, $E_i$ and $E_j$, on shell, 
the bare coupling constants are removed by defining a renormalized amplitude:
\begin{equation}
\mathcal{A}_{ij}^R(\mathbf{p},\mathbf{k},E)=\sqrt{|Z_i||Z_j|}\mathcal{A}_{ij}(\mathbf{p},\mathbf{k},E) \:,
\end{equation}
where $Z_i$ is the residue of the dimer pole $i$ in the vacuum,
\begin{equation}
Z_i=\frac{32 \pi}{g_i^2 a_i} \:.
\label{Res-faktoren}
\end{equation}
This  renormalized amplitude has the same poles in the three-body sector as the 
three-particle scattering amplitude. We now expand the fermion-dimer amplitude in partial waves as
\begin{equation}
\mathcal{A}_{ij}^R(\mathbf{p},\mathbf{k},E) = \sum_{l=0}^\infty (2l+1)\,(\mathcal{A}^R_{ij})_l[p,k,E]
P_l(\cos\theta_k)\:,
\label{partial waves}
\end{equation}
where $\cos \theta_k = \mathbf{p}\cdot \mathbf{k}/(pk)$ and $P_l$ is a Legendre 
polynomial. The different partial waves decouple and the integral equation for
the $l$th partial wave amplitudes is
\begin{align}
\label{eq:ampliudeeq}
i(\mathcal{A}^R_{ij})_l[p,k,E] =& \frac{1}{2} \frac{-8\pi i}{\sqrt{|a_i||a_j|}} \int_{-1}^1 d 
\cos \theta_k P_l( \cos \theta_k) 
\, t_{ij}(p,k, \theta_k,E) \notag \\
&+  i\sum_{r=0}^2 4 \pi \frac{\sqrt{|a_r|}}{\sqrt{|a_i|}} \int_{k_F}^{\Lambda} \frac{dq}{(2\pi)^2}q^2 
\int_{-1}^1 d\cos \theta_q \, P_l(\cos \theta_q) \,t_{ir}(p,q,\theta_q,E) \,    \notag \\
& \times  \overline{D}_r (q,E) \mathcal({A}_{rj}^R)_l[q,k,E] \:, 
\end{align}
where 
\begin{align}
t_{ij}(p,k,\theta_{k},E):=&\frac{\theta (|\mathbf{p}+\mathbf{k}|-k_F)(1-\delta_{ij})}{E-p^2 -k^2 -pk
\cos  \theta_{k} + i \epsilon} \:,
\end{align}
and
\begin{equation}
\overline{D}_r(q,E):= \left[\frac{1}{a_r}-\frac{1}{\pi}L(E-\einhalb q^2,q)\right]^{-1} \:
\end{equation}
is the dimer propagator without prefactors. In the vacuum only the S-wave amplitude has bound state
poles. This remains true in the medium and we thus focus on the poles of the S-wave in-medium
amplitude $(\mathcal{A}^R_{ij})_0[p,k,E]$. The technical details of the implementation of the boundary
conditions from the Pauli blocking are discussed in Appendix \ref{sec:medium-integ3}. In the next
section, we present our results for the pole structure of $(\mathcal{A}^R_{ij})_0[p,k,E]$.

\subsection{Results}
\begin{figure}[t]
\centering
\includegraphics[clip,width=8cm,angle=0]{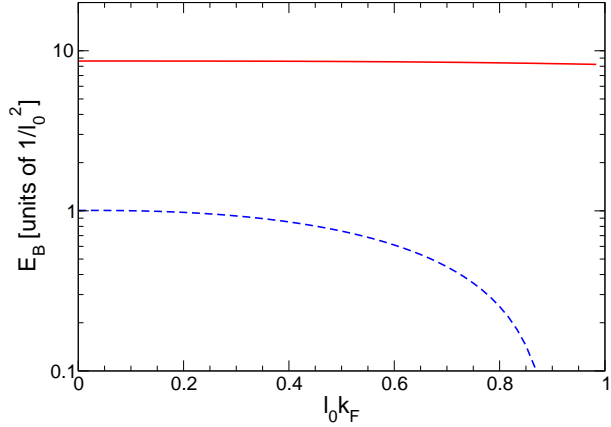}
\caption{(Color online) 
Binding energy $E_B$ of two states depicted in dependence of $k_F$:
$a_k=l_0$ for $k=0,1,2$ and $\Lambda=250/l_0$.}
\label{fig:spektrum}
\end{figure}

In this section we discuss the poles  of the the S-wave amplitude $(\mathcal{A}^R_{ij})_0$, 
in general for three different scattering lengths.
In Eq.~(\ref{eq:ampliudeeq}) the three-body force dependence was traded for the cutoff dependence, 
so that the cutoff determines the three-body energy in the vacuum for given scattering lengths.
A spectrum of two states as a function of the Fermi momentum is shown in Fig. \ref{fig:spektrum}. 
Similar to the two-body case, the binding energy of each state decreases with rising 
Fermi momentum due to medium effects. Remarkable is the difference of the energy loss with increasing Fermi 
momentum for shallow and deep states. The  
less bound state disappears through the threshold while the more deeply bound state
looses only about 5\% of its binding energy as $k_F l_0$ is increased from 0 to 1.
This behaviour of the 
three-body spectra is generic and was always observed in our calculations.

\begin{figure}[t]
\centering
\includegraphics[clip,width=12cm]{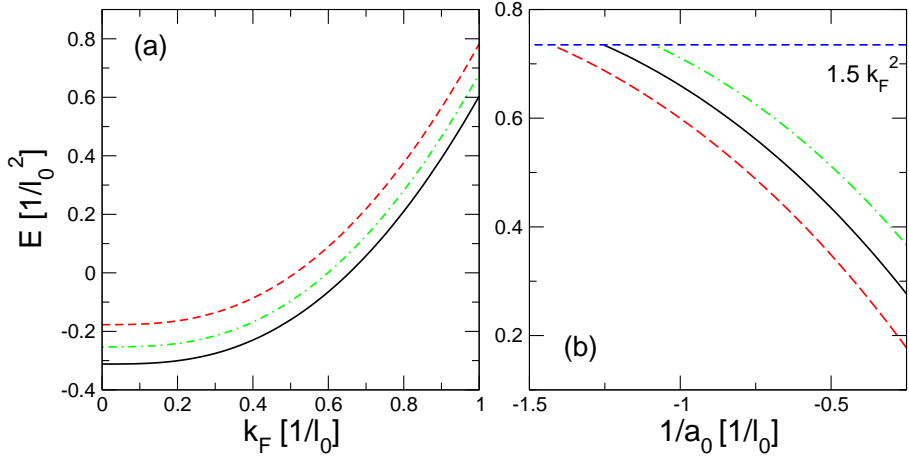}
\caption{(Color online)
(a) Energy of three-body poles plotted against $k_F$ for $a_0=-1.0l_0$ (dashed), $a_0=-1.25l_0$ 
(dash-dotted) and $a_0=-1.5l_0$ (solid); $\Lambda=150/l_0$ and the remaining scattering lengths are 
$a_1=-2l_0$ and $a_2=-2.5 l_0$.
(b) The pole energy is shown as a function of $1/a_0$ for   $\Lambda= 150/l_0$ (dash-dotted), $\Lambda=160/l_0$ 
(solid) and $\Lambda= 170/l_0$ (dashed); $k_F=0.7/l_0$ and the remaining scattering lengths are 
$a_1=-l_0$ and $a_2=-0.99l_0$. The horizontal line gives $E=1.5 \, k_F^2$.}
\label{fig:coopertripel}
\end{figure}
In Fig. \ref{fig:coopertripel}~(a) the energy 
of a generic three-body pole is plotted against the Fermi momentum for three negative scattering lengths. 
As in the previous case, the binding energy reduces with increasing  Fermi momentum. Indeed, 
the energy goes to zero and  continuously rises to positive values. Hence, we have found poles with 
positive energy. Since the total  momentum is  zero, they can not correspond to bound states.
Note the resemblance between this figure and Fig.~\ref{fig:dimer-kf2} which shows dimer poles. 

To get a better understanding of  these positive energy poles,  we have 
varied one of  the three negative scattering lengths while keeping the other two constant, see 
Fig. \ref{fig:coopertripel}~(b). 
The energy rises with decreasing $1/a$, but vanishes when 
the value of the energy becomes  $1.5 \,k_F^2$. For different configurations of the Fermi momenta, 
scattering lengths, and the cutoff, we have always found this threshold. The accuracy of the location
of this threshold reaches  to the third  (fourth) decimal place for cutoffs of the order 100 (10) $l_0$.
In order to explain this observation, we draw an analogy with the positive energy poles in the two-body case. 
There, the energy gain $\Delta E$ is the kinetic energy minus the energy of the pole. Hence, the energy of 
the pole can not be larger than the maximum kinetic energy. 
\begin{figure}[ht]
\centering
\includegraphics[clip,width=10cm,angle=0]{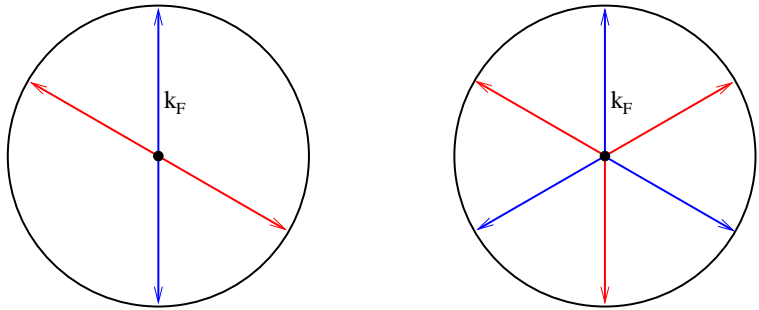}
\caption{(Color online)
Configurations in the Fermi sphere for two (left) and three particles (right) with total momentum  $P=0$.}
\label{fig:skizze-fermikugel}
\end{figure}
In Fig.~\ref{fig:skizze-fermikugel} configurations of two and three particles inside the Fermi 
sphere are shown for total momentum $P=0$. As discussed in the previous section, 
the maximum two-body pole energy is $k_F^2$. 
In the case of three particles the magnitude of each momentum can be 
$k_F$, whereas the total momentum remains zero. So the maximum kinetic  energy of three particles 
inside the Fermi sphere is $3 \times k_F^2/2=1.5 k_F^2$. 
We conjecture that the  three-particle poles 
belong to a state similiar to a Cooper pair, but built out of three particles, which we call a \lq\lq Cooper triple''. 
In contrast to Cooper pairs, these Cooper triples are fermions. 
If the three pair scattering lengths are equal, the Cooper triples
are SU(3) singlets. However, for different scattering lengths, the 
SU(3) symmetry is broken.

Cooper triples also appear if one scattering length is
positive. Since the energy of the triples is continuous in $1/a_i$ ($i=0,1,2$), the region
of three negative scattering lengths merges into the region of one positive
and two negative scattering lengths at the point $1/a_i=0$ (the other two
scattering lengths are considered constant). Therefore the pole energy has
to remain positive in the limit $1/a_i \rightarrow 0^-$ to obtain Cooper
triples for one positive scattering length. For this scenario, the Fermi
momentum must be sufficiently large. The actual value depends on the two constant scattering
lengths. Hence, Cooper triples also occur in this region. An analogous argument holds if two
or three scattering lengths are positive. In all three cases, we have observed Cooper triples
in our calculations.
However, it remains to be verified that the Fermi spheres assumed in
our calculation persist in this region.

\begin{figure}[t]
\centering
\includegraphics[clip,width=8cm,angle=0]{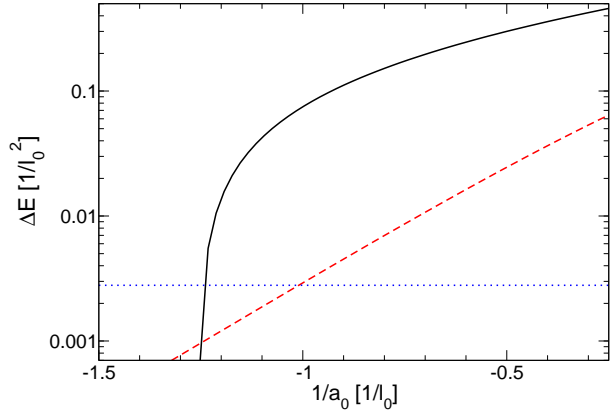}
\caption{(Color online)
$\Delta E$ is plotted against the inverse scattering length $1/a_0$ of three-body (solid line) and 
two-body (dashed line) poles; $\Lambda=160/l_0$, $k_F=0.7/l_0$ and the constant scattering lengths are
$a_1=-l_0$ and $a_2=-0.99l_0$. The 
horizontal dotted line shows $\Delta E$ for the constant scattering length $a_2=-0.99l_0$. 
}
\label{fig:deltaE}
\end{figure}
Next, we examine which state is energetically favorable. If Cooper pairs are built in a three component 
Fermi gas, the pairs are typically formed between two components while the residual component remains unpaired. 
Therefore, we compare the energy gain of a Cooper triple, $1.5 \,k_F^2$ 
minus pole energy, with the energy gain of a Cooper pair, $k_F^2$ minus the pole energy.
The energy gain $\Delta E$ of a Cooper pair and a Cooper triple are compared  as a function of one 
variable scattering length in Fig. \ref{fig:deltaE}. The remaining parameters stay the same as in 
Fig.~\ref{fig:coopertripel}~(b). Since $\Delta E$ of the Cooper pair depends on the scattering length, 
it can be energetically favorable to build a different Cooper pair connected to one of the constant scattering 
lengths. To account for this, we have also plotted the energy gain of the larger constant scattering length. 
We find that the energy gain of the three-particle poles is much larger (note the logarithmic axis). Only near the 
threshold for the triple, the three-particle $\Delta E$ rapidly falls off and drops below the energy gain of both Cooper  
pairs. This suggests that Cooper triples could play an important role in three-component 
Fermi gases in the continuum.

In principle, it should always be possible to find these positive energy poles for three negative scattering 
lengths. In contrast to the two-body case, the poles do not newly emerge in the medium. Primarily, 
the poles were bound states in the 
vacuum which became modified by the medium, see Fig.~\ref{fig:coopertripel}~(a). Thus, the Fermi momentum
must be large enough to obtain positive energy poles. This is most easily achieved for
Cooper triples emerging from rather shallow three-body bound states in vacuum.

\section{Conclusion and Outlook}
In this paper, we have examined the influence of Pauli blocking on universal two- and three-body states. 
First the poles of the two-body scattering amplitude in the medium were regarded. We were able to
recover the physics of Cooper pairs and BEC-BCS crossover from the pole structure of the amplitude. 
In particular, we found that the binding energy of bound states decreases with rising  Fermi 
momentum due to medium effects. 
In the negative scattering length region, positive energy poles emerge which can be 
identified with Cooper pairs. 
In the crossover region, the poles show a different behaviour and their nature changes 
with the total momentum. They can not be uniquely identified as bound states or 
Cooper pairs.

We have used the same strategy to investigate the pole structure of the three-body
scattering amplitude. We found that the medium effects reduce the binding
of three-body states compared to the vacuum. This is in agreement with the findings
of Ref.~\cite{Nygaard:2011aa}, where the modification of the Efimov
spectrum for three equal-mass fermions with one of the fermions
embedded in a Fermi sea was calculated. Moreover, we found
three-body poles with positive energy. As in the two-body sector,
we observed a continuous crossover from 
negative energy poles to positive energy poles as the Fermi momentum is varied. 
The maximum energy of the poles was found to be $1.5 k_F^2$. In analogy to 
the connection between positive energy poles in the two-body sector and 
Cooper pairs, we 
have interpreted
this as evidence for the formation of Cooper triples 
composed out of three particles.
These Cooper triples are fermions and thus can not Bose condense.
The energy gain of such a triple was found to be larger than the energy gain of the corresponding 
Cooper pair over a large region of scattering lengths. Consequently, it appears to be energetically favorable 
to form a triple instead of a pair and an unpaired third atom in this region.

In the case of equal pair scattering lengths, the Cooper triples are 
SU(3) singlets. For different scattering lengths, however, the SU(3) 
symmetry is broken (cf. Fig.~\ref{fig:coopertripel}). 
If the three scattering lengths are large, the 
SU(3) breaking is small since the leading corrections to the SU(3) limit are
proportional to the inverse scattering lengths \cite{Mehen:1999qs}.

How these three-body correlations affect a many-body system is an open question.
It would be interesting to extend previous studies of the phase structure
of three-component Fermi gases 
\cite{pair-three-comp,super-phases,Breached-Pairing} to include the triples
and investigate their influence.
A qualitative picture of the many-body
structure in the SU(3) symmetric limit was given 
by Floerchinger and collaborators \cite{Floerchinger:2009fp}. 
They argue that at small density, if the scattering length is varied from
large negative to large positive values, the BCS and BEC phases are separated 
by a trion phase of three-body bound states. At small densities, our Cooper 
triples must reduce to the trions of Ref.~\cite{Floerchinger:2009fp}.
A related study for three-component fermions in an optical lattice was 
carried out in Ref.~\cite{Rapp:2006rx}. 
Within a Hubbard model with SU(3) symmetry, a trion 
phase of three-fermion bound states 
has been predicted at strong coupling and a parallel to the 
baryonic phase of QCD was drawn.

There may also be a connection to Ref.~\cite{Schuck-BF}, where Boson-Fermion 
(BF) interactions were regarded in a similiar analysis. BF pairs 
at positive energies were found and the ground state was assumed to be a 
Fermi Gas of BF-Cooper pairs, since the pairs are still fermions. The 
interaction of three distinguishable fermions in our case could also be 
regarded as the interaction of a Cooper pair (Boson) and  a  fermion of the 
remaining type, if the scattering lengths are negative (BCS region) and at 
least two scattering lengths are different. In this case, we can conjecture 
that the ground state of the system is  a Fermi gas of Cooper triples, which 
are composites of a Cooper pair and a unpaired fermion.

Since we have only included the Pauli blocking effects from the medium, 
further theoretical study 
is required. This could for example be achieved by performing Monte 
Carlo simulations of such
systems similar to the two-flavour case \cite{Lee:2008fa}. Such a 
calculation would allow for more quantitative
predictions of the effect. In analogy to Ref.~\cite{Schuck-BF}, 
the triples might also lead to a new type
of superfluidity in three-component Fermi systems which 
could be observed in ultracold atomic gases.
For this purpose, it would be useful to calculate the interactions 
of the triples. 
If their interactions are attractive, they could again form Cooper
pairs and Bose condense.
Much insight would be gained if one could calculate 
the energy of such a condensate 
and compare it with a BCS condensate. This would allow to determine
under which conditions such a new type of superfluidity might occur.
An experimental test of this scenario could be carried out
with $^6$Li atoms where mixtures of three different hyperfine
states with tunable interactions are already available
\cite{Ottenstein:2008,Huckans:2008,Lompe:10,Nakajima:11}.

\begin{acknowledgments}
We thank R.J.~Furnstahl for discussions and Dean Lee for comments on the manuscript.
Partial financial support from the Deutsche Forschungsgemeinschaft (SFB/TR 16),
and BMBF~(grant 06BN9006) are acknowledged.  This work was further supported by the 
EU HadronPhysics3 project ``Study of strongly interacting matter''.
\end{acknowledgments}

\appendix

\section{Medium integrals}
\label{sec:medium-integ}
This section gives some details of the calculation of the loop integral for the full in-medium dimer-propagator. 
The loop integral $I(P_0,\mathbf{P})$ is defined as follows 
\begin{equation}
iI(P_0,\mathbf{P})  = \vierint \frac{i\Theta \left(| \frac{ \mathbf{P}}{2}  +\mathbf{q}| 
-k_F \right) }{\frac{P_0}{2}+q_0 -\frac{1}{2}\left( \frac{\mathbf{P}}{2} +\mathbf{q} \right)^2 +i\epsilon}     
\frac{i\Theta \left(| \frac{ \mathbf{P}}{2}  -\mathbf{q}| -k_F \right) }{\frac{P_0}{2}-q_0 
-\frac{1}{2}\left( \frac{\mathbf{P}}{2} -\mathbf{q} \right)^2 +i \epsilon}
\:.
\end{equation}
After a contour integration the integral simplifies to
\begin{equation}
iI(P_0,\mathbf{P}) = i \dreiint \frac{\Theta \left(| \frac{ \mathbf{P}}{2}  +\mathbf{q}| 
-k_F \right) \Theta \left(| \frac{ \mathbf{P}}{2}  -\mathbf{q}| -k_F \right)}{P_0-\frac{P^2}{4}-q^2+i\epsilon } \:.
\end{equation}
As already mentioned, the theta functions are boundary conditions to the integral. We choose $\mathbf{P}$ 
to be aligned in $z$-direction and switch to spherical coordinates. The $\phi$ integration still gives $2 \pi$. 
But the lower $q$ boundary depends on $\theta$, the angle between $\mathbf{P}$ and $\mathbf{q}$. The two theta 
functions are equivalent to the following conditions:
\begin{align}
f^{-}(q):=q^2-Pqx+\tfrac{1}{4}P^2 -k_F^2>0 \:,\\
f^{+}(q):=q^2+Pqx+\tfrac{1}{4}P^2-k_F^2>0   \:,
\end{align}
where $x=\cos \,\theta= \mathbf{P}  \mathbf{q}/(Pq)$. The functions $f^{\pm}(q)$ are simple parabolas, whose roots are
\begin{align}
&f^{+}: \quad -\frac{Px}{2} \pm \sqrt{\frac{P^2}{4}(x^2-1)+k_F^2} \:,\\
&f^{-}: \quad +\frac{Px}{2} \pm \sqrt{\frac{P^2}{4}(x^2-1)+k_F^2} \:.
\end{align}
These roots do only exist for all  $x \in [-1,1]$ if $\frac{P}{2}\leq k_F$. Therefore we have to 
distinguish the two cases $P>2k_F$ and $P<2k_F$.
\begin{itemize}
\item[(a)] $P<2k_F$:
\\The theta functions move the lower boundary $a(x)$ which is

 \begin{equation}
a(x)= \begin{cases}
\frac{Px}{2}+\sqrt{ \frac{P^2}{4}(x^2-1)+k_F^2 } & \text{for } x>0 \\
-\frac{Px}{2}+\sqrt{ \frac{P^2}{4}(x^2-1)+k_F^2 } & \text{for } x<0
\end{cases} \:.
\end{equation}
The upper boundary remains unchanged. After a rescaling: $ \frac{P}{2}=s \cdot k_F$ , $ q=t \cdot k_F $, $\widetilde{\Lambda} 
=\Lambda/k_F$ and $ b = \left( P_0 - \frac{P^2}{4} + i \epsilon \right)/k_F^2 $, the integral can be written as
\begin{align}
iI(P_0,\mathbf{P})=&\frac{i k_F}{(2\pi)^2} \Bigg\{ \int_0^1 \int^{\tilde{\Lambda}}_{sx+\sqrt{s^2(x^2-1)+1}} \,\frac{t^2}{b-t^2} \,dt dx  \\
&+ \int_{-1}^0 \int^{\tilde{\Lambda}}_{-sx+\sqrt{s^2(x^2-1)+1}} \, \frac{t^2}{b-t^2} \,dt dx \Bigg\} \:. 
\end{align}
One can easily see that the second integral merges to the first if the substitution $x \rightarrow -x$ is performed.
\item[(b)] $P>2k_F$: 
\\In this case the $q$ integration range is
\begin{equation}
\begin{cases}
\left[0,\frac{Px}{2}-\sqrt{c(x)}\right] \mbox{ and }\left[ \frac{Px}{2}+\sqrt{c(x)},\Lambda\right] &\mbox{ for }
 \sqrt{1-\frac{1}{s^2}}<x \leq 1 \notag  \\
\left[0,\Lambda \right]& \mbox{ for }  -\sqrt{1-\frac{1}{s^2}} <x<\sqrt{1-\frac{1}{s^2}} \notag \\
\left[ 0,-\frac{Px}{2}-\sqrt{c(x)}\right] \mbox{ and } \left[ -\frac{Px}{2}+\sqrt{c(x)},\Lambda\right] &
\mbox{ for } -1 \leq x < -\sqrt{1-\frac{1}{s^2}} 
\end{cases} \:,
\end{equation}
with $\sqrt{c(x)} := \sqrt{ k_F^2 -\frac{P^2}{4}(1-x^2) } $.
Therefore the integral becomes
\begin{align}
iI(P_0,P)=& \frac{ik_F}{(2\pi)^2} \Bigg \{ \int_{-1}^{-\sqrt{1-\frac{1}{s^2}}} \left(\int_0^{-sx-\sqrt{c'(x)}} \frac{t^2}{b-t^2}dt +\int_{-sx+\sqrt{c'(x)}}^{\tilde{\Lambda}} \frac{t^2}{b-t^2}dt \right)dx \notag \\
&+   \int_{-\sqrt{1-\frac{1}{s^2}}}^{\sqrt{1-\frac{1}{s^2}}} \int_0^{\tilde{\Lambda}}  \frac{t^2}{b-t^2} \,dtdx \notag\\ &+\int_{\sqrt{1-\frac{1}{s^2}}}^1 \left(\int_0^{sx-\sqrt{c'(x)}} \frac{t^2}{b-t^2} dt +\int_{sx+\sqrt{c'(x)}}^{\tilde{\Lambda}} \frac{t^2}{b-t^2}dt\right)dx \Bigg \} \:,
\end{align}
with $c'(x)=1-s^2(1-x^2)$. The integrals in the first and third line are equal, similiar to the preceding case.
\end{itemize}

\section{Integral kernel}
\label{sec:medium-integ3}
In this section, we discuss the calculation of the integral
\begin{equation}
\int_{-1}^1 d\cos \theta_q \, P_l(\cos \theta_q) \,t_{ik}(p,q,\theta_q,E)  \:,
\end{equation}
with
\begin{align}
t_{ij}(p,q,\theta_{q},E):=&\frac{\theta (|\mathbf{p}+\mathbf{q}|-k_F)(1-\delta_{ij})}{E-p^2 -q^2 -pq
\cos  \theta_{q} + i \epsilon} \:,
\end{align}
which is required to derive the integral equation for the three-body amplitudes in
the medium.
This type of integral appears in the inhomgeneous as well as in the homgeneous part. 
The theta function is a boundary condition on the $cos \,\theta_q$ integration:
\begin{equation}
\theta( |\mathbf{p}+\mathbf{q}|-k_F) \Rightarrow p^2 +2pq \cos \theta_q + q^2 >k_F^2\:.
\end{equation}
Two cases have to be distinguished. First, if $|p-q|$ is larger than  $k_F$, the theta function is 
always fulfilled. Consequently, the integration region is $[-1,1]$. If $|p-q|<k_F$, the lower boundary 
will be changed. Note that the $q$ integration begins at $k_F$. The lower boundary $\theta_g$ is
\begin{equation}
\cos \theta_g = \frac{k_F^2 -p^2 -q^2}{2pq} \:.
\end{equation}
The angle integration for the S-wave can now be written
\begin{align}
&\int_a^1 dx\frac{P_0(x)}{E-p^2-q^2 -pqx + i \epsilon} =\frac{1}{pq}\int_a^1 dx \, \frac{1}{c-x} = -\frac{1}{pq} \left[ \ln(c-x) \right]_a^1\:,\notag \\
\end{align}
with  $c=\frac{1}{pq} (E-p^2-q^2+ i \epsilon)$. 
The result is\\\\
$|p-q|>k_F$ : $a=-1$\\
\begin{align}
 -\frac{1}{pq} \left[ \ln(c-x) \right]_a^1=& \frac{1}{pq}\left( \ln \left( \frac{E-p^2-q^2+pq+ i \epsilon}{pq}\right) - \ln \left(\frac{E-p^2-q^2-pq + i \epsilon}{pq} \right) \right)\:,
\end{align}
$|p-q|<k_F$ : $a=\cos \,\theta_g$\\
\begin{align}
 -\frac{1}{pq} \left[ \ln(c-x) \right]_a^1 = & \frac{1}{pq}\left( \ln \left( \frac{E - \einhalb p^2 - \einhalb q^2 - \einhalb k_F^2 + i \epsilon}{pq}\right)-\ln \left( \frac{E-p^2-q^2-pq + i \epsilon}{pq} \right) \right)\:.
\end{align}


\begin{thebibliography}{99}

\bibitem{Giorgini:2008}
S. Giorgini, L.P. Pitaevskii, and S. Stringari,
Rev.\ Mod.\ Phys.\ {\bf 80} (2008) 1215
[arXiv:0706.3360 [cond-mat.other]].

\bibitem{Ketterle:2008}
W. Ketterle, M. Zwierlein, 
in 
Proceedings of the International School of 
Physics \lq\lq Enrico Fermi'', 
Course CLXIV, edited by M. Inguscio, W. Ketterle, 
and C. Salomon (IOS Press, Amsterdam, 2008)
[arXiv:0801.2500 [cond-mat.other]]. 

\bibitem{Lee:2008fa}
D. Lee,
Prog.\ Part.\ Nucl.\ Phys. {\bf 63} (2009) 117
[arXiv:0804.3501 [nucl-th]].

\bibitem{Chin:2010aa}
C. Chin, R. Grimm, P. Julienne, and E. Tiesinga,
Rev.\ Mod.\ Phys. {\bf 82} (2010) 1225
[arXiv:0812.1496 [cond-mat.other]].

\bibitem{Tan:2005xx}
S. Tan,
Ann.\ Phys.\  {\bf 323} (2008) 2952 [arXiv:cond-mat/0505200],
ibid. 2971 [arXiv:cond-mat/0508320],
ibid. 2987 [arXiv:0803.0841 [cond-mat.stat-mech]].

\bibitem{Braaten:2010if}
  E.~Braaten,
  Lect.\ Notes Phys.\  {\bf 836} (2012) 193
  [arXiv:1008.2922 [cond-mat.quant-gas]].

\bibitem{Efimov:1970zz}
  V.~Efimov,
  Phys.\ Lett.\  {\bf B33 } (1970)  563.


\bibitem{Ottenstein:2008}
  T. B. Ottenstein, T. Lompe, M. Kohnen, A. N. Wenz, and S. Jochim,
  Phys.\ Rev.\ Lett.\ {\bf 101} (2008) 203202
  [arXiv:0806.0587 [cond-mat.other]].

\bibitem{Huckans:2008}
  J.H.~Huckans, J.R.~Williams, E.L.~Hazlett, R.W.~Stites, and K.M.~O'Hara,
  Phys.\ Rev.\ Lett.\ {\bf 102} (2009) 165302
  [arXiv:0810.3288  [physics.atom-ph]].

\bibitem{Braaten:2008wd}
  E.~Braaten, H.-W.~Hammer, D.~Kang and L.~Platter,
  Phys.\ Rev.\ Lett.\  {\bf 103} (2009) 073202
  [arXiv:0811.3578 [cond-mat.other]].

\bibitem{schmidt:2008fz}
  S.~Floerchinger, R. Schmidt and C.~Wetterich,
  Phys.\ Rev.\  A {\bf 79} (2009) 053633
  [arxiv:0812.1191 [cond-mat.other]].

\bibitem{Naidon:2009}
  P.~Naidon, M.~Ueda,
  Phys.\ Rev.\  Lett.\ {\bf 103} (2009) 073203.

\bibitem{Braaten:2009ey}
  E.~Braaten, H.-W.~Hammer, D.~Kang and L.~Platter,
  Phys.\ Rev.\  A {\bf 81} (2010) 013605
 [arXiv:0908.4046 [cond-mat.quant-gas]].


\bibitem{Lompe:10}
  T.\ Lompe, T.B.\ Ottenstein, F.\ Serwane, A.N.\ Wenz, G.\ Z\"urn,
  and S.\ Jochim,
  Science {\bf 330} (2010) 940 
  [arXiv:1006.2241 [cond-mat.quant-gas]].

\bibitem{Nakajima:11}
  S.\ Nakajima, M.\ Horikoshi, T.\ Mukaiyama, P.\ Naidon, and M.\ Ueda,
  Phys.\ Rev.\ Lett.\  {\bf 106} (2011) 143201 
  [arXiv:1010.1954 [cond-mat.quant-gas]].


\bibitem{pair-three-comp} T.~Paananen, J.-P.~Martikainen, and P.~T\"orm\"a, 
Phys. Rev. A {\bf 73} (2006) 053606 
[arXiv:cond-mat/0603498 [cond-mat.supr-con]].

\bibitem{super-phases} P.F.~Bedaque and J.P.~D'Incao,
Annals Phys. {\bf 324} (2009) 1763 
[arXiv:cond-mat/0602525 [cond-mat.other]].

\bibitem{Breached-Pairing} B.~Errea, J.~Dukelsky, and G.~Ortiz, 
Phys.\ Rev.\  A {\bf 79} (2009) 051603 
[arXiv:0812.2395 [cond-mat.supr-con]].

\bibitem{Catelani:2008}
G.~Catelani and E.A.~Yuzbashyan, Phys.\ Rev.\ A {\bf 78}, 033615 (2008)
[arXiv:0805.3663].

\bibitem{Crossover-3komp} T.~Ozawa and G.~Baym, 
Phys.\ Rev.\ A {\bf 82} (2010) 063615
[arXiv:1011.0467 [cond-mat.quant-gas]].

\bibitem{Rapp:2006rx}
  A.~Rapp, G.~Zarand, C.~Honerkamp and W.~Hofstetter,
  Phys.\ Rev.\ Lett.\  {\bf 98} (2007) 160405
  [cond-mat/0607138].

\bibitem{P-Block-Ef} D.J.~MacNeill and F.~Zhou, 
Phys.\ Rev.\ Lett.\ {\bf 106} (2011) 145301
[arXiv:1011.0006 [cond-mat.quant-gas]].

\bibitem{Nygaard:2011aa}
N. G. Nygaard and  N. T. Zinner,
arXiv:1110.5854 [cond-mat.quant-gas].

\bibitem{Schuck-BF}
A.~Storozhenko, P.~Schuck, T.~Suzuki, H.~Yabu, and J.~Dukelsky,
Phys.\ Rev.\ A {\bf 71} (2005) 063617;
X.~Barillier-Pertuisel, S.~Pittel, L.~Pollet, and P.~Schuck,
Phys.\ Rev.\ A {\bf 77} (2008) 012115;
T.~Watanabe, T.~Suzuki, and P.~Schuck,
Phys.\ Rev.\ A {\bf 78} (2008) 033601.

\bibitem{Hammer:2000xg}
  H.-W.~Hammer and R.~J.~Furnstahl,
  Nucl.\ Phys.\ A {\bf 678} (2000) 277
  [nucl-th/0004043].


\bibitem{Furnstahl:2008df}
  R.~J.~Furnstahl, G.~Rupak and T.~Schafer,
  Ann.\ Rev.\ Nucl.\ Part.\ Sci.\  {\bf 58} (2008) 1
  [arXiv:0801.0729 [nucl-th]].


\bibitem{Steele:2000qt}
  J.~V.~Steele,
  [nucl-th/0010066].

\bibitem{Schafer:2005kg}
  T.~Schafer, C.~-W.~Kao and S.~R.~Cotanch,
  Nucl.\ Phys.\ A {\bf 762} (2005) 82
  [nucl-th/0504088].

\bibitem{Hammer:2000nf}
  H.-W.~Hammer, T.~Mehen,
  Nucl.\ Phys.\  {\bf A690 } (2001)  535.
  [nucl-th/0011024].

\bibitem{hw-review} E. Braaten and H.-W. Hammer, 
Phys.\ Rept.\  {\bf 428} (2006) 259   
[arXiv:cond-mat/0410417 [cond-mat.other]].

\bibitem{Fetter} A.L. Fetter and J.D. Walecka, 
\emph{Quantum Theory of Many-Particle Systems} 
(Dover Publications, 2003).

\bibitem{Braaten:2002sr}
  E.~Braaten, H.-W.~Hammer and M.~Kusunoki,
  Phys.\ Rev.\ A {\bf 67} (2003) 022505
  [cond-mat/0201281].

\bibitem{Bedaque:1998kg}
  P.~F.~Bedaque, H.-W.~Hammer, U.~van Kolck,
  Phys.\ Rev.\ Lett.\  {\bf 82 } (1999)  463
  [nucl-th/9809025].

\bibitem{PatrickDip}
Patrick Niemann, ``Wenigteilcheneffekte im Medium,'' Diploma thesis,  Universit\"at Bonn (2010).

\bibitem{Mehen:1999qs}
  T.~Mehen, I.~W.~Stewart and M.~B.~Wise,
  Phys.\ Rev.\ Lett.\  {\bf 83} (1999) 931
  [hep-ph/9902370].

\bibitem{Floerchinger:2009fp}
  S.~Floerchinger, R.~Schmidt, S.~Moroz and C.~Wetterich,
  Phys.\ Rev.\ A {\bf 79} (2009) 013603
  [arXiv:0809.1675 [cond-mat.supr-con]].

\end{thebibliography}
\end{document}